\documentclass[11pt,a4paper]{article}
\usepackage{jcappub}
\usepackage[english]{babel}
\usepackage[utf8]{inputenc}

\usepackage[dvipsnames]{xcolor}
\usepackage{url}
\usepackage{graphicx}
\usepackage{color}
\usepackage{ulem}
\usepackage{booktabs}
\usepackage{hyperref}
\usepackage{placeins}
\usepackage{caption}
\usepackage{subcaption}
\usepackage{bm}
\usepackage{siunitx}

\numberwithin{equation}{section}

\newcommand{\GeV}{\mbox{GeV}}

\newcommand{\mdm}{m_{\chi}}
\newcommand{\Fermi}{\textit{Fermi}}

\newcommand{\sixs}{\sigma_{\mathrm{SI}}}
\newcommand{\sv}{\langle\sigma_{\textrm{ann}} v \rangle}

\newcommand{\params}{{\mathbf \Theta}}
\newcommand{\like}{\mathcal{L}}

\dedicated{UCI-HEP-TR-2015-21}

\title{Effective Field Theory of Dark Matter: a Global Analysis}

\author[a]{Sebastian Liem,}
\author[a]{Gianfranco Bertone,}
\author[a]{Francesca Calore,}
\author[b]{Roberto Ruiz de Austri,}
\author[c]{Tim M.P. Tait,}
\author[d,e]{Roberto Trotta,}
\author[a]{Christoph Weniger}

\affiliation[a]{GRAPPA, University of Amsterdam, Science Park 904, 1098 XH Amsterdam, Netherlands}
\affiliation[b]{Instituto de F\'isica Corpuscular, IFIC-UV/CSIC, Valencia, Spain}
\affiliation[c]{Department of Physics and Astronomy, University of California, Irvine, CA 92697, USA}
\affiliation[d]{Imperial College London, Astrophysics \& Imperial Centre for Inference and Cosmology,
Blackett Laboratory,  SW7 2AZ London, United Kingdom}
\affiliation[e]{Imperial College London, Data Science Institute, William Penney Laboratory,   SW7 2AZ London, United Kingdom}

\emailAdd{sebastian.liem@uva.nl}

\abstract{We present global fits of an effective field theory description of real, and complex scalar dark matter candidates. We simultaneously take into account all possible dimension 6 operators consisting of dark matter bilinears and gauge invariant combinations of quark and gluon fields. We derive constraints on the free model parameters for both the real (five parameters) and complex (seven) scalar dark matter models obtained by combining Planck data on the cosmic microwave background, direct detection limits from LUX, and indirect detection limits from the \Fermi~Large Area Telescope. We find that for real scalars indirect dark matter searches disfavour a dark matter particle mass below 100 GeV. For the complex scalar dark matter particle current data have a limited impact due to the presence of operators that lead to p-wave annihilation, and also do not contribute to the spin-independent scattering cross-section. Although current data are not informative enough to strongly constrain the theory parameter space, we demonstrate the power of our formalism to reconstruct the theoretical parameters compatible with an actual dark matter detection, by assuming that the excess of gamma rays observed by the \Fermi~Large Area Telescope towards the Galactic centre is entirely due to dark matter annihilations. Please note that the excess can very well be due to astrophysical sources such as millisecond pulsars.
We find that scalar dark matter interacting via effective field theory operators can in principle explain the Galactic centre excess, but that such interpretation is in strong tension with the non-detection of gamma rays from dwarf galaxies in the real scalar case. In the complex scalar case there is enough freedom to relieve the tension.}

\begin{document}

\maketitle

\flushbottom


\section{Introduction}
Overwhelming observational evidence points to  the existence of dark matter (DM)
, from Galactic up to cosmological scales
~\cite{Bertone:2010zza,Jungman:1995df,Bergstrom00,Bertone05}
The most widely discussed and well-motivated 
particle DM candidates are weakly interacting massive particles (WIMP), which arise from some of the most popular extensions of the SM, and naturally lead to the right DM abundance~\cite{Bertone05}.

Well-motivated ultraviolet-complete particle-physics models accommodate good DM particle candidates in their mass spectrum. This is the case
for example of supersymmetric theory with R-parity conservation~\cite{Jungman:1995df} or Universal Extra Dimensions theories~\cite{Servant:2002aq}. On the other hand, 
one can adopt a model independent approach that makes minimal assumptions on the DM particle and its couplings with SM particles. In the framework 
of Effective Field Theories (EFT) for particle DM, DM would be the only additional degree of freedom beyond the SM accessible by
current experiments~\cite{Beltran:2010ww}. Therefore, the interactions of the DM particle with SM particles are described by effective operators (of dimension 6 or higher).
Those can be predictive if the energy scale of the experiment under investigation is lower than the energy scale of the operator's coefficients, while 
will break down once the energy scale of the experiment is of the order of the mass of any particle mediating the DM -- SM  interactions.

Several detection strategies are applied in order to detect the elusive nature of DM. Direct detection searches look for the recoil energy of nuclei scattered off by DM particles in large, underground laboratories. Indirect detection searches aim to detect the final stable products (like gamma rays, neutrinos or charged cosmic rays) of DM annihilation or decay above the large astrophysical background. Finally, searches for DM at colliders are based on the possibility to look for the production of new particles beyond the SM.

EFT have been shown to be able to capture the main features of generic WIMP candidates in their range of validity and being a
powerful framework to compare theory against data.  
EFT operators have been studied extensively in context of various experiments, see for example \cite{Goodman:2010ku, Goodman:2010yf, Goodman:2010qn}.
However, in these studies each operator was considered separately to draw phenomenological implications and detectability prospects. 
In this work, we present the first global analysis of \emph{all} relevant EFT operators simultaneously in light of the latest constraints from indirect and direct searches for DM.  
We perform a Bayesian statistical scan of the EFT parameter space, as described in section~\ref{s:eftop}, by implementing all the latest experimental constraints, following an approach similar to that adopted for supersymmetric scenarios in Refs. \cite{Trotta:2008bp,Roszkowski:2009sm,Feroz:2011bj,Bertone:2011nj,Strege:2011pk,Strege:2012bt,Strege:2014ija,Bertone:2015tza}.
Previously, Bal\'azs et al. \cite{Balazs:2014rsa} presented a similar work with a full set of EFT operators. Our analysis goes beyond that of Ref. \cite{Balazs:2014rsa} in several respects: we perform a more thorough statistical analysis that addresses the dependence on priors, and we perform a comparison between profile likelihood and posterior distributions; we include the contribution from 
DM--gluon operators that was previously neglected while being potentially sizeable; and we make use of the latest data from the \Fermi~Large Area Telescope (\Fermi-LAT).
We do not include collider constraints from the LHC, 
as these do not rigorously exist for scalar dark matter, and are typically subdominant to direct search
constraints for the masses of particular interest to us here \cite{ATLAS:2012ky,Khachatryan:2014rra}.

This work is organised as follows. In section \ref{s:eftop} we describe the theory of EFT operators we use. In section \ref{s:method} we describe our statistical approach and computational method, in section \ref{s:results} we present the results, and in section \ref{s:conclusions} we present our conclusions.


\section{Effective field theory operators}\label{s:eftop}

We focus on the interactions of DM with quarks and gluons, since these interactions are currently probed very efficiently by so-called `direct detection' experiments, and by searches for DM production at the LHC.  It is relatively
straightforward to extend our formalism to include leptons (e.g. \cite{Fox:2011fx}), but we leave such refinements for future work.  
We further make some well-motivated simplifying assumptions
about the nature of the interactions.  In particular,
we work in the limit in which the particles mediating the interactions between the DM, quarks, and gluons are heavy compared to the energies
of interest.  In that limit, all theories map into an EFT which encapsulates the interactions of the DM with the Standard Model via a set
of non-renormalizable interactions \cite{Beltran:2008xg,Beltran:2010ww,Goodman:2010yf,Bai:2010hh,Goodman:2010ku},
\begin{equation}
\mathcal{L} \supset \sum_i \lambda_i ~ \mathcal{O}_i,
\end{equation}
where the coefficients $\lambda_i$ have dimensions of inverse mass to the appropriate power such that the over-all dimension of $\mathcal{L}$ remains four
and the $\mathcal{O}_i$ are a set of operators consisting of a DM bilinear contracted with a gauge invariant combination of quark and/or gluon
fields.

\begin{table}
\centering
\begin{tabular}{@{}cllcc@{}}
\toprule
\multicolumn{5}{c}{Real scalar DM operators} \\
\midrule
Label & Coefficient & Operator & $\sixs$ & $\sv$\\
\midrule
R1 & $\lambda_1 \sim \frac{1}{2 M^2}$ & $ m_q  \chi^2 \bar{q} q$ & \checkmark & s-wave  \\
R2 & $\lambda_2 \sim \frac{1}{2 M^2} $ & $i m_q  \chi^2 \bar{q} \gamma^5 q$ & & s-wave  \\
R3 & $\lambda_3 \sim \frac{\alpha_s}{4 M^2}$ & $\chi^2 G_{\mu\nu}G^{\mu\nu}$ & \checkmark & s-wave  \\
R4 & $\lambda_4 \sim \frac{\alpha_s}{4 M^2}$ & $i \chi^2 G_{\mu\nu}\tilde{G}^{\mu\nu}$ & & s-wave  \\

\bottomrule
\multicolumn{5}{c}{Complex scalar DM operators} \\
\midrule
Label & Coefficient & Operator & $\sixs$ &  $\sv$\\
\midrule
C1 & $\lambda_1 \sim \frac{1}{M^2}$ & $ m_q \chi^\dag \chi  \bar{q} q$ & \checkmark & s-wave \\
C2 & $\lambda_2 \sim \frac{1}{M^2}$ & $ i m_q \chi^\dag \chi  \bar{q} \gamma^5 q$ & & s-wave  \\
C3 & $\lambda_3 \sim \frac{1}{M^2}$ & $  \chi^\dag \partial_\mu \chi \bar{q} \gamma^\mu q$ & \checkmark & p-wave \\
C4 & $\lambda_4 \sim \frac{1}{M^2}$ & $ \chi^\dag \partial_\mu \chi \bar{q} \gamma^\mu \gamma^5 q$ & & p-wave \\
C5 & $\lambda_5 \sim \frac{\alpha_s}{8 M^2}$ & $\chi^\dag \chi G_{\mu\nu}G^{\mu\nu}$ & \checkmark &  s-wave \\
C6 & $\lambda_6 \sim \frac{\alpha_s}{8 M^2}$ & $i \chi^\dag \chi  G_{\mu\nu}\tilde{G}^{\mu\nu}$ & & s-wave \\
\bottomrule
\end{tabular}
\caption{The EFT operators for real and complex scalar DM interacting with quarks and gluons.  Also indicated is the mapping 
for each coefficient $\lambda_i$ to the notation of \cite{Goodman:2010ku}, which
operators contributes to $\sigma_{\rm SI}$, and if $\langle \sigma_{\rm ann} v \rangle$ is s-wave or p-wave dominated in the non-relativistic limit. \label{t:operators}}
\end{table}

As a starting point, we focus on the case in which the DM is a single species of (real or complex) scalar particle that is a singlet under
the electroweak symmetry, as this limits the number
of Lorentz structures describing its interactions\footnote{Our formalism also applies rather simply to the case of fermionic DM, which requires more parameters to 
describe its interactions. \cite{Goodman:2010ku}}.  We restrict our basis of operators to those which are leading in the sense of being the least marginal at low
energies and which represent the leading structures consistent with the principle of minimal flavor violation (MFV) \cite{D'Ambrosio:2002ex}, which dictates that
their contributions to flavour-changing neutral currents follows the same CKM structure of the SM itself, mitigating the otherwise extreme constraints
from the null searches for non-SM sources of flavour violation.  After rotating the quarks into the mass basis, 
this effectively results in the quark vector bilinears having a generation-independent coupling whereas the scalar bilinears are weighted by the quark mass.
The resulting set of operators for real (R) and complex (C) DM are shown in Table~\ref{t:operators}.  Also indicated in the table are the
operators which make velocity-unsuppressed contributions 
to spin-independent scattering with nuclei ($\sigma_{\rm SI}$) or annihilation ($\langle \sigma_{\rm ann} v \rangle$)
in the non-relativistic limit.
Note that the possible annihilation channels for these models are the kinematically available quarks and gluons. 
The branching ratios are determined by the relative strength of the operators.
For each individual operator the ratios to different flavours (when applicable) are democratic except for the operators weighted by the quark mass.
For these the branching ratios to different quark flavours do have an $m_q^2$ dependence, i.e. the heaviest available is favoured.

While typically an analysis will assume that one operator or another dominates, any realistic UV model of scalar DM will involve several in concert
with related coefficients (see \cite{Bai:2010hh,Gershtein:2008bf,Chang:2013oia,An:2013xka,Bai:2013iqa,DiFranzo:2013vra,Buckley:2014fba,Baek:2015lna} for examples).  
To truly represent the heavy-mediator limit in general, one must allow for combinations of interactions.  Combined with the DM mass, this defines
a parameter set of five quantities for real DM and seven for complex.  This work represents the first truly general analysis of the scalar singlet
DM parameter space in the EFT limit.

It is also worth mentioning that the EFT description will fail to accurately describe observables whose typical momentum transfer is large enough to be
on the order of the particles mediating the interaction.  Particularly for large momentum transfer processes such as at the LHC, this implies that limits derived
in an EFT context do not apply to models in which the mediator masses are $\lesssim$~TeV \cite{Abdallah:2015ter,Abercrombie:2015wmb}.

\FloatBarrier

\newpage
\section{Statistical framework}\label{s:method}

We use highly efficient Bayesian methods to explore the models but we present our results both in Bayesian and in frequentist terms.
Our approach is based on Bayes' theorem (see e.g. ~\cite{Trotta:2008qt})

\begin{equation}
p(\params|\mathbf{D})=\frac{p(\mathbf{D}|\params) p(\params)}{p(\mathbf{D})},
\label{eqn:Bayes}
\end{equation}
where $\mathbf{D}$ are the data and $\params$ are the model parameters of interest. Bayes' theorem states that the posterior probability distribution function (pdf) $p(\params|\mathbf{D})$ for the parameters is obtained from the likelihood function $p(\mathbf{D}|\params) \equiv \like(\params)$ and the prior pdf (or ``prior'' for short) $p(\params)$. In this article we are primarily interested in parameter inference, therefore the Bayesian evidence $p(\mathbf{D})$ merely act as a normalisation constant, and will not be considered further in the following analysis. 

In order to study the constraints on a single parameter of interest $\theta_i$, one can consider either the one-dimensional marginal posterior, or the one-dimensional profile likelihood. The marginal posterior is obtained from the full posterior distribution by integrating (marginalising) over the unwanted parameters in the $n$-dimensional parameter space:

\begin{equation}
p(\theta_i|\mathbf{D})=\int p(\params|\mathbf{D}) d\theta_1 ... d\theta_{i-1} d\theta_{i+1} ... d\theta_{n}.
\end{equation}

On the other hand, the profile likelihood function for $\theta_i$, instead, is found by maximising over the parameters that are not of interest:

\begin{equation}
{\mathcal L}(\theta_i)= \max_{\theta_1,...,\theta_{i-1},\theta_{i+1},...,\theta_{n}}\mathcal{L}(\params).		
 \end{equation}

The extension of these concepts to more than one parameter is straightforward.
The profile likelihood and the marginal posterior are two different statistical quantities that may lead to different conclusions about the parameter space of interest.
The marginal posterior integrates over hidden parameter directions and therefore correctly accounts for volume effects; it peaks at the region of highest posterior mass.
The profile likelihood peaks at the region of highest likelihood.
It is oblivious to volume effects, but is an excellent quantity to find small regions of high likelihood in parameter space. 
These two quantities do not necessarily lead to the same conclusions for non-Gaussian likelihoods, and the maximum of information about the model parameter space is obtained by studying both of these quantities.
Therefore, in the following we present results for both the marginalised Bayesian posterior and the profile likelihood.


\subsection{Experimental constraints and the likelihood function}
The experimental constraints are implemented as a joint likelihood function $\mathcal{L}$ with each component representing different contraints.

\begin{equation}
\ln \mathcal{L} = \ln \mathcal{L}_{\Omega_{\chi}h^2} + \ln \mathcal{L}_{\textrm{DD}} + \ln \mathcal{L}_{\textrm{CMB}} + \ln  \mathcal{L}_{\textrm{dSph}} + \ln  \mathcal{L}_{\textrm{GCE}},
\end{equation}
where  $\mathcal{L}_{\Omega_{\chi}h^2}$ is the part corresponding to measurements of the cosmological DM relic density, $\mathcal{L}_{\textrm{DD}}$  direct DM detection constraints and 
$\mathcal{L}_{\textrm{CMB}}, \mathcal{L}_{\textrm{dSph}}, \mathcal{L}_{\textrm{GCE}}$ are from DM indirect detection constraints. 
We discuss each component in turn:

\paragraph{$\ln \mathcal{L}_{\Omega_{\chi}h^2}$:} We apply a Gaussian likelihood taken the Planck cosmic microwave background (CMB) data constraint on the DM relic abundance. 
We use as central value the result from Planck temperature and lensing data $\Omega_{\chi}h^2 = 0.1186 \pm 0.0031$ \cite{Ade:2013zuv} with a (fixed) theoretical uncertainty, 
$\tau = 0.012$, to account for the numerical uncertainties entering in the calculation of the relic density. 

\paragraph{$\ln \mathcal{L}_{\textrm{DD}}$:} For DM direct detection we use
upper limits from the LUX experiment \cite{Akerib:2013tjd}, as implemented in
the \texttt{LUXCalc} code~\cite{Savage:2015xta}. We adopt  
hadronic matrix elements determined by lattice QCD \cite{QCDSF:2011aa,Junnarkar:2013ac}. 
We use a local DM density of $\rho_{\mathrm{dm}} = 0.4$ GeV cm$^{-3}$.

\paragraph{$\ln \mathcal{L}_{\textrm{CMB}}$:} DM annihilating into ionising particles during the cosmic dark ages will broaden the last scattering surface of the CMB. 
This would modify the CMB anisotropies as measured by the Planck satellite.
As Planck has not measure any such effect \cite{Ade:2015xua} we can use the data put limits on our models.
We use the likelihood as defined by \cite{Cline:2013fm} using the updated analysis from \cite{Slatyer:2015jla}. 

\paragraph{$\ln  \mathcal{L}_{\textrm{dSph}}$:} As limits from indirect searches
for DM, we focus here on limits from gamma ray observations with the \Fermi-LAT.
As the arguably most robust limits in indirect
searches, we adopt constraints that were obtained from the non-observation of a
gamma ray signal from a dozen of dwarf spheroidal galaxies.  We use here
results from \cite{Ackermann:2015zua}, which are based on a combined analysis
of six years of \Fermi-LAT data, and take into account uncertainties in the DM
content of each dwarf spheroidal.  The results of that analysis were presented
as tabulated likelihood functions, which allows us to apply them to models with
arbitrary gamma ray spectra.  

\paragraph{$\ln  \mathcal{L}_{\textrm{GCE}}$:} Given the excitement about a possible
gamma ray DM signal from the Galactic centre (see \cite{Daylan:2014rsa, TheFermi-LAT:2015kwa, Huang:2015rlu} and
references therein), we also include a likelihood that evaluates the
compatibility between this gamma ray excess and a DM annihilation signal as
predicted by our models.  To this end, we use the results from
\cite{Calore:2014xka}, which account for systematic correlated uncertainties
related to the subtraction of Galactic diffuse foregrounds along the
line-of-sight towards the Galactic centre.

\smallskip

Both the dwarf spheroidal and the
Galactic centre likelihoods we have used are conveniently packaged in the
\texttt{gamLike}\footnote{C.~Weniger et al., to be released soon.} code.

\subsection{Priors}\label{s:priors}

As we are using Bayesian inference we need to provide prior probability distributions for our parameters. For the DM mass parameter $\mdm$  we consider the typical range $\mdm \in [1,1000] \: \GeV$ advocated for WIMPs, and we adopt a prior uniform on the log of the quantity, which reflecs a state of indifference with regards to the mass scale.

Given a mediator particle with mass $M$ with couplings $g_1, g_2$ to the DM and SM particles respectively, we can write the operator coefficients $\lambda_i$ as

\begin{equation}
\lambda_i = k_i \frac{g_1 g_2}{M^2} \, , 
\end{equation}
\noindent
where $k_i$ is an operator specific constant. In this way, conditions on the underlying physics can then translate onto the coefficients.
The theory has to be perturbative $g_1 g_2 \leq (4\pi)^2$, and, for the EFT description to be valid, we, at least, need $M > 2 \mdm$. These conditions imply

\begin{equation}
\lambda_i < k_i \frac{4\pi^2}{\mdm^2} \, .
\end{equation}

The bound's strong dependence on $\mdm$ makes the operator coefficients very hard to scan efficiently.
We could define hyperparameters to eliminate the mass dependence, but together with the unavoidable lower bound, this would introduce a prior preference towards higher DM masses.
Instead we use the fact that, as we shall see later, the constraint from producing the correct relic density is much more constraining than the theoretical bound.
We therefore simply impose $\lambda_i < 1$. 
There is no physical motivation for a lower bound, so we conservatively adopt $10^{-20}$, since if $\lambda_i = 10^{-20}$ then the operators $\mathcal{O}_i$ do not contribute significantly to any observable.

In order to assess the prior dependence of the posterior, we consider here two different kind of priors to scan over $\lambda_i$. We first consider a prior uniform on the log, i.e. $\log_{10} \lambda_i \in [-20, 0]$. Then, we define two hyperparameters:  a common energy scale to all operators $A$, and the coefficients $f_i$ such that $\lambda_i = f_i A$. For $A$ we use a prior uniform on $\log_{10} A \in [-20, 0]$.
For $f_i$ we use a symmetric Dirichlet prior between 0 and 1. The $f_i$ determine if the operator $\mathcal{O}_i$ contributes or not; we impose $\sum f_i = 1$ to ensure that at least one operator contributes, and that the energy scale is actually determined only by $A$.
The symmetric Dirichlet distributions are parametrised by a single scalar $\alpha$ which determines how the $f_i$'s are distributed.
Larger values lead all $f_i$ to be similar, while smaller values tend to select a few large $f_i$. For the prior we use we have $\alpha = 0.1$.  
Later when we refer to these two priors as the Log and Dirichlet prior respectively.

\subsection{Scanning methodology}

To map out the posterior and the likelihood we have developed the \texttt{EFTBayeS} package in which we implemented the Lagrangians of our models in \texttt{FeynRules} \cite{Alloul:2013bka}.
This is then interfaced with a modified version of \texttt{MicrOMEGAs} v2.4 \cite{Belanger:2006is} to compute the relic abundance of DM and direct and indirect detection rates.
The code \texttt{LUXCalc} v1.0.1 \cite{Savage:2015xta} has been used to compute the LUX experiment likelihood and \texttt{PPPC 4 DM ID} \cite{Cirelli:2010xx, Ciafaloni:2010ti} tables to calculate the photon spectrum of DM annihilation products.
We have used tabulated values of $f_{\text{eff}}$ from \cite{Slatyer:2015jla}. $f_{\text{eff}}$ is the efficiency for DM annihilations to deposit energy into the gas medium in the cosmic dark ages and is used in the CMB constraint.

For the exploration of the EFT models, the \texttt{EFTBayeS} code uses MultiNest v2.18~\cite{Feroz:2007kg,Feroz:2008xx} nested sampling algorithm. MultiNest  is an extremely efficient scanning algorithm that can reduce the number of likelihood evaluations required for an accurate mapping of the posterior pdf by up to two orders of magnitude with respect to conventional MCMC methods. This Bayesian algorithm, originally designed to compute the model likelihood and to accurately map out the posterior, is also able to reliably evaluate the profile likelihood, given appropriate MultiNest settings, as demonstrated in~\cite{Feroz:2011bj}.


\newpage
\section{Results}\label{s:results}

In this section we present the impact of aforementioned experimental data on the parameter space of the real and the complex scalar DM candidates. 
We first show the profile likelihood and posteriors when applying the relic density determination and all current experimental limits.
We then show what happens when we also assume that the Galactic centre excess is entirely due to the annihilation of our DM candidates.
This will serve as an example to demonstrate the constraining power of an astroparticle detection, when combined with the relic density constraint.


\subsection{Impact of Planck, LUX, and \Fermi-LAT} \label{s:2d}

\begin{figure}
  \centering
     \includegraphics[width=0.95\textwidth]{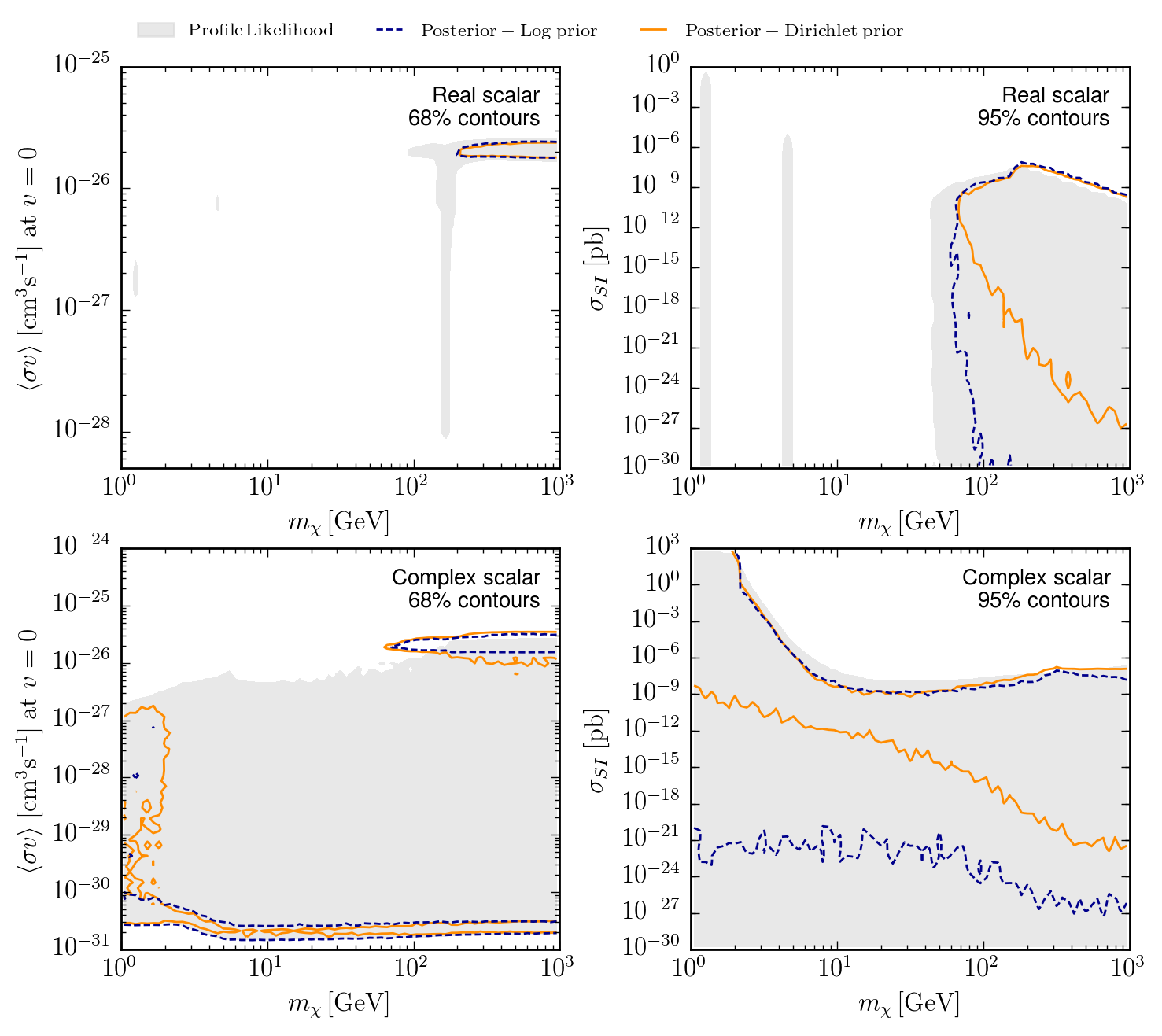}
  \caption{Confidence region and credibility regions for real and complex scalar DM candidates in observable vs mass planes.
  In the $\mdm, \sv$ panels we show the 68\% contours while in the $\mdm, \sixs$ panels we show the 95\% contours.
  This is done for clarity of presentation; the posteriors in certain regions are fairly flat which leads to very noisy contours.
  The experimental constraints applied are the relic density of DM, limit on spin-independent scattering cross-section from LUX, and annihilation cross-section limits from both the Planck CMB measurement, and the stacked dSph analysis from \Fermi-LAT.
  The different posteriors are using different priors; the Log prior and the Dirichlet prior ($\alpha = 0.1$) as defined in section \ref{s:priors}.
  \label{f:2D}}
\end{figure}

In figure \ref{f:2D} we show the two 2D marginalised posterior and the 2D profile likelihood for both the real and complex scalar DM candidate in the planes of $\mdm$ and either $\sv$ or $\sixs$.
This is after including all experimental limits in the likelihood, but excluding the Galactic centre excess.
In order to illustrate the dependence of the results on priors, we show the posterior distribution under the Log and the Dirichlet priors as defined in section \ref{s:priors}.

We begin with discussing the features of in the $\sv$ panels.
We see that the value of the annihilation cross-section is quite well-constrained, due to the fact that it is directly related to the relic density, which is a well measured quantity.
For the complex scalar candidate we have two regions in the posteriors, because the C3 and the C4 operators are p-wave rather than s-wave as the rest of the operators are (all real scalar operators are s-wave.) 
In the lower region it is one of the p-wave operators that dominates, while in the upper region it is one of the s-wave operators.
By a dominating operator we mean that the corresponding coefficient $\lambda_i$ is large enough to provide the correct amount of relic density by itself.
The differences we see between the posteriors and the profile likelihood is due to the difference between marginalisation (i.e. integration) and profiling (i.e. maximation).

In the complex scalar case, the values of the annihilation cross-section between the p-wave region and the s-wave region still have a large profile likelihood as there is at least one combination of input parameters for which the operator coefficients give rise to the correct relic density.
However, as there are not very many of these possible combinations in the input parameter space, they have a negligible effect on the posteriors.

In the real scalar case the most noticeable difference between the profile likelihood and the posteriors are the downward spikes in $\sv$ when the DM mass is close to a quark mass.
This is a kinematical effect that enhances the cross-section in the early universe when new annihilation channels open up \cite{Griest:1990kh}.
In this case the new channels are additional quarks in the R1/C1, R2/C2 operators. 
This effect isn't present in today's colder universe which is why we get the downward spikes in $\sv$ today.
These are tiny, i.e. highly tuned, regions in the parameter spaces which is why we only see them in the profile likelihood.

The bottom of the two spikes at $\mdm \approx 1.3$ GeV and $4.7$ GeV (i.e. charm and bottom quark masses) avoid all limits.
In fact the only term in the likelihood that has any impact  for $\mdm < 5$ GeV is the CMB term, whereas the constraining power of the
LUX experiment vanishes as the recoil energies of these low mass DM candidates fall below the detection threshold.
As for \Fermi-LAT dSph results, they should have an impact on these region, but the lowest photon energy ($E_{\gamma}$) considered in the likelihood published by \Fermi-LAT and implement here, is 500 MeV, and for $\mdm \lesssim 5$ GeV the spectra actually peaks below $E_{\gamma} = 500$ MeV.
For $\mdm = 5$ GeV 75\% of the photons has an energy less than 500 MeV, for $\mdm = 1$ GeV that number increases to 99\%.\footnote{
These numbers do not significantly depend on the annihilation channel in question.
The spectra are quite similar for all quarks and gluons.}
We strongly suspect that these low mass regions would be disfavoured in an analysis extended to lower $E_{\gamma}$.

This limitation in the dSph likelihood also explains the existence of slightly favoured regions seen in the complex scalar profile likelihood maps for $\mdm$ and $\sv$ in figure \ref{f:C_lim} below.
There is a slight upward fluctuation in the low energy bins of the dSph likelihood, this does not matter for larger $\mdm$ as their spectra is fitted against a large set of bins.
However, as stated above, for low $\mdm$ much of the spectra is below the lowest energy bin of 500 MeV and the importance of the slight upward fluctuation is enhanced.

In contrast with the annihilation cross-section the spin-independent scattering cross-section is only bounded from above by the relic density.
This is because not all operators contribute to $\sixs$, which means that as long as one of the non-contributing operators' coefficient is large enough to provide the correct relic density, the coefficients of the contributing operators are free to be arbitrarily low.
This means that the profile likelihood region actually extends down to arbitrarily low numbers.
This freedom to put the operator coefficients at arbitrary low values is also the reason why the two posteriors differ, i.e. the tails of the distributions depend on the choice of priors for the operator coefficients.

\begin{figure}
\centering
\includegraphics[width=0.90\textwidth]{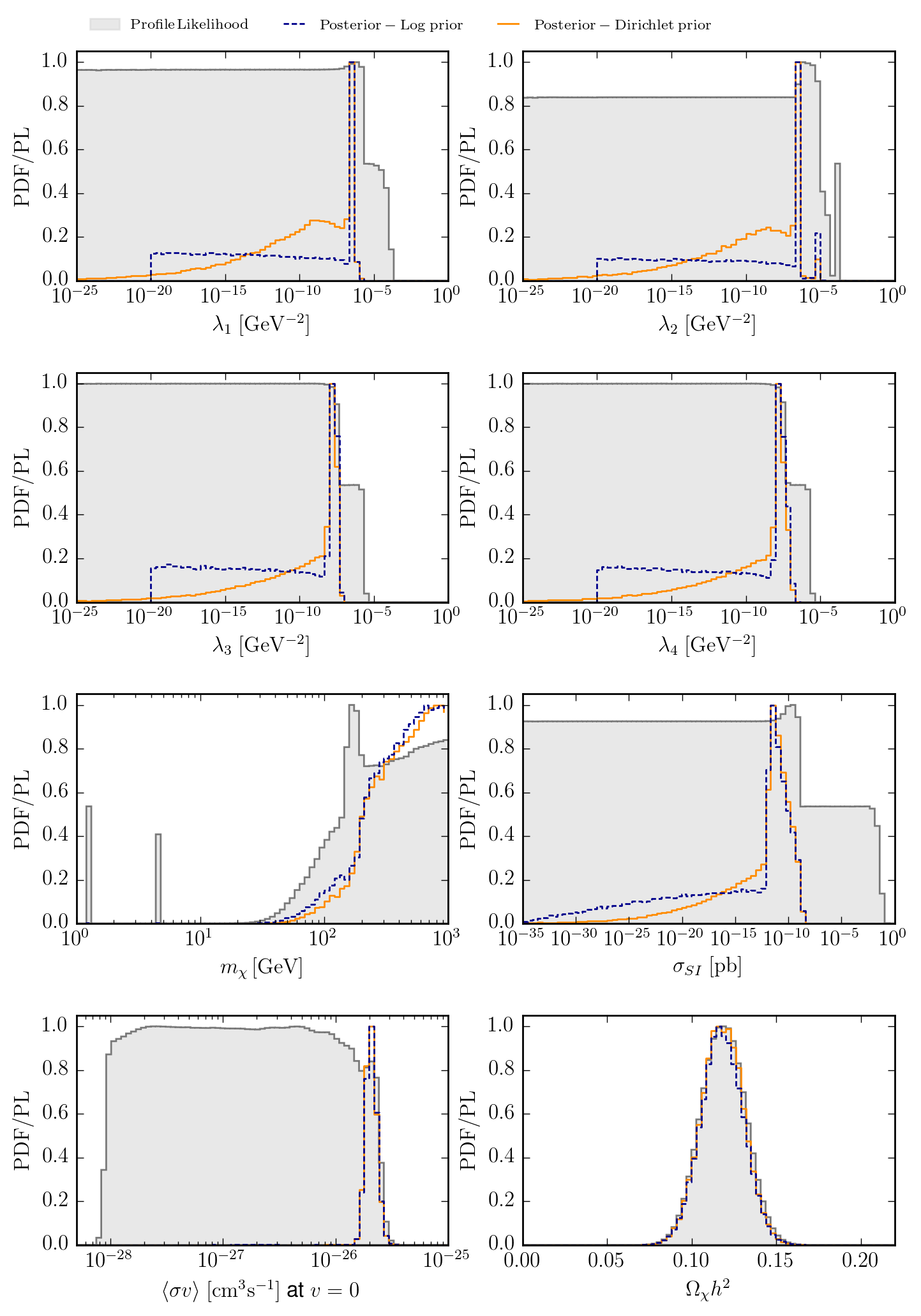}
\caption{1D posterior distributions and profile likelihood of parameters and observables for real scalar DM.
  The experimental constraints considered are the relic density of DM, limit on spin-independent scattering cross-section from LUX, and annihilation cross-section limits from both the Planck CMB measurement, and the stacked dSph analysis from \Fermi-LAT.
  The different posteriors are using different priors; the Log prior and the Dirichlet prior ($\alpha = 0.1$) as defined in section \ref{s:priors}.
Posteriors and likelihoods are normalised to their peaks. 
 \label{f:R_lim}}
\end{figure}

This is easier to understand if we look at the 1D posteriors and profile likelihood in figure \ref{f:R_lim} and \ref{f:C_lim} for real and complex scalar respectively. 
In the panels for the operator coefficient $\lambda_i$ we clearly see that the tails are prior dependent. The profile likelihood is in fact flat for low $\lambda_i$ and following Bayes' theorem \eqref{eqn:Bayes}, when the likelihood is constant, then the posterior simply traces the prior.
We note however that the position of the peaks does not depend on the choice of prior (although the height does, since it depends on the size of the tails through the normalisation of the posterior) and they tell us that the most probable solution is when a single operator provides all the DM relic density by themselves.

For the real scalar candidate low masses are effectively disfavoured by Planck, \Fermi-LAT and LUX with the exception of the small regions around $1.3$ GeV and $4.7$ GeV as discussed earlier. 
In the complex scalar case the mass is much less constrained due to the presence of p-wave operators.

\subsection{GCE as example of additional measurement}

\begin{figure}
\centering
\includegraphics[width=0.90\textwidth]{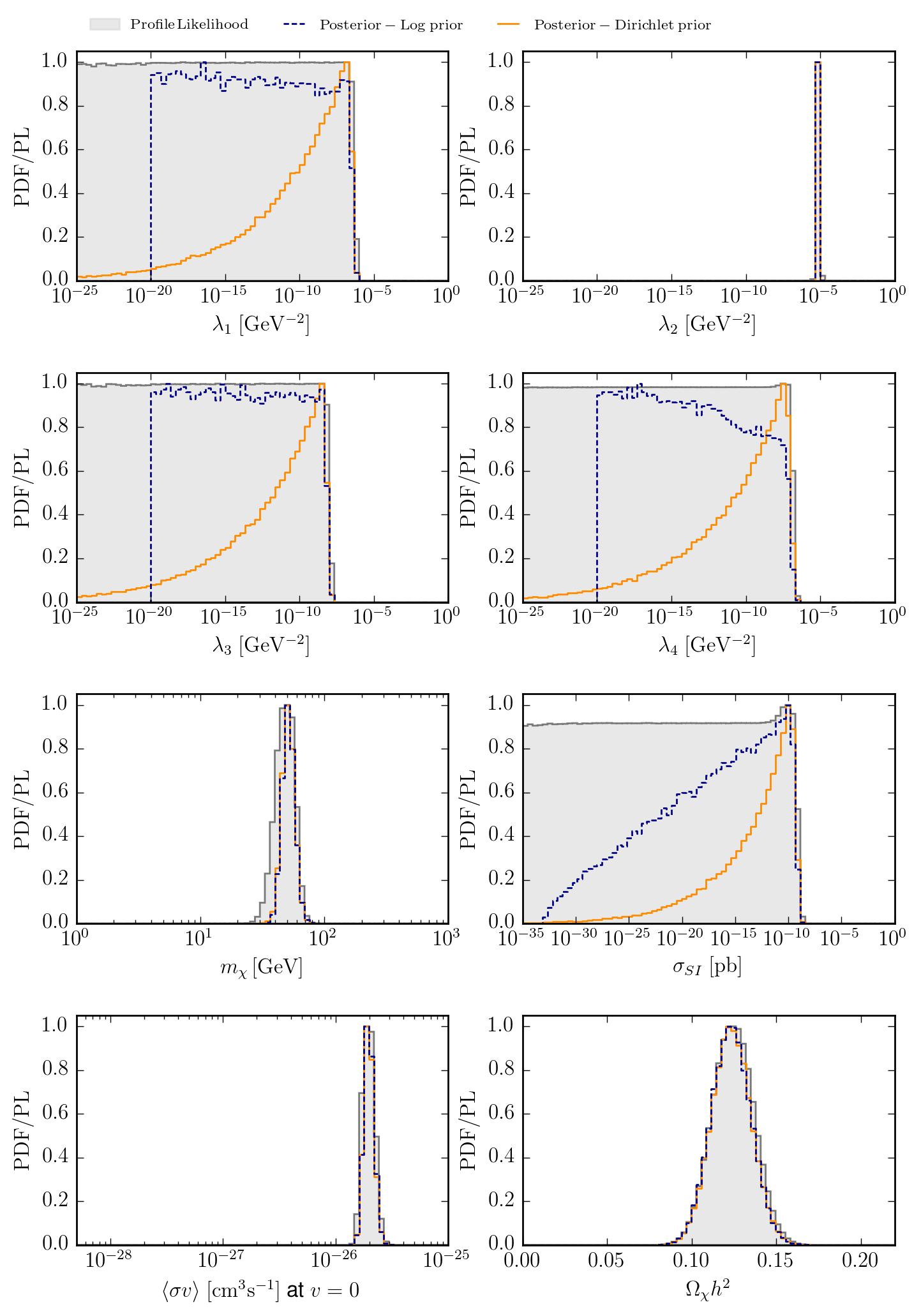}
\caption{1D posterior distributions and profile likelihood of parameters and observables for real scalar DM when assuming it explains the Galactic centre excess.
  The additional experimental constraints considered are the relic density of DM, limit on spin-independent scattering cross-section from LUX, and annihilation cross-section limits from both the Planck CMB measurement, and the stacked dSph analysis from \Fermi-LAT.
  The different posteriors are using different priors; the Log prior and the Dirichlet prior ($\alpha = 0.1$) as defined in section \ref{s:priors}.
  Posteriors and likelihoods are normalised to their peaks.
\label{f:R_gce}}
\end{figure}

\begin{figure}
\centering
\includegraphics[width=0.90\textwidth]{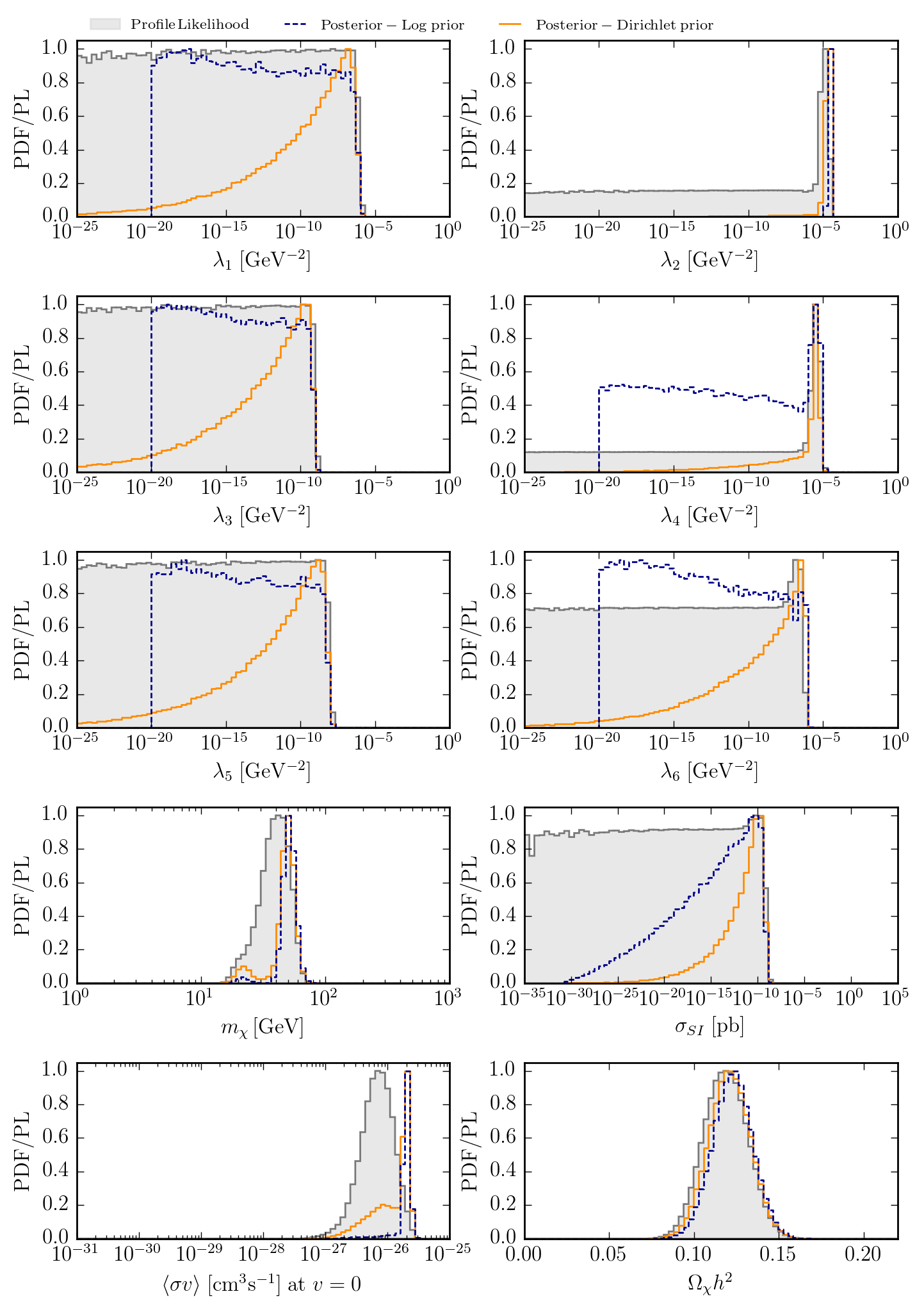}
\caption{1D posterior distributions and profile likelihood of parameters and observables for complex scalar DM when assuming it explains the Galactic centre excess.
  The additional experimental constraints considered are the relic density of DM, limit on spin-independent scattering cross-section from LUX, and annihilation cross-section limits from both the Planck CMB measurement, and the stacked dSph analysis from \Fermi-LAT.
  The different posteriors are using different priors; the Log prior and the Dirichlet prior ($\alpha = 0.1$) as defined in section \ref{s:priors}.
 Posteriors and likelihoods are normalised to their peaks.
\label{f:C_gce}}
\end{figure}

In figure \ref{f:R_gce} and \ref{f:C_gce} we show the 1D profile likelihood and posteriors for real and complex scalar DM candidate when we include an actual measurement in the likelihood, in addition to the relic density.
The measurement we consider is the Galactic centre excess because it is an actual well-studied measurement, and still compatible with a DM interpretation. 
Although a more standard astrophysical interpretation, e.g. in terms of millisecond pulsars \cite{Bartels:2015aea, Yuan20141, PhysRevD.90.023526, PhysRevD.88.083521, Abazajian:2010zy,PhysRevLett.116.051103}, is possible or even probable, we assume here that the excess is fully explained by our DM candidates.

The impact of including this measurement in the likelihood is drastic. For the real scalar case only the R2 operator is the possible solution with a $\mdm \approx$ 40--60 GeV, whereas the R1 and R3 operators are disfavoured because they contribute to the spin-independent scattering cross-section and are therefore disfavoured by LUX; and the R4 operator favours slightly lower masses which are excluded by the \Fermi-LAT dSph analysis.

In fact, even the R2 operator is in tension with the limit from the dSph analysis, but it is the least disfavoured operator. 
If we look closer at the best fit point from the scan in table \ref{t:bfpoints} we see that it is a good fit to the excess. 
However, if we compare with the best fit point when not fitting the excess we see that the other data, mainly the dSph, contributes a $\Delta \chi^2 = 5.1$ which is sizeable.
In summary, if one believes the Galactic centre excess is due to DM annihilation, the tension with the non-detection of gamma rays from dwarf galaxies would disfavour the real scalar model is excluded as the dark matter candidate. 

In the case of the complex scalar candidate, the two additional operators, C3 and C4, are p-wave operators and thus cannot explain the excess in themselves.
They do, however, relax the connection between the relic density and the annihilation cross-section which somewhat relieves the tension from the dSph limit.
If we look at the best fit point for the complex case in table \ref{t:bfpoints} we again see that it is a good fit to the excess.
Comparing with the best fit point without the excess we see a $\Delta \chi^2 = 1.78$. 

However, the freedom in $\sv$ also means we can increase the Galactic centre J-factor in order to lower the annihilation cross-section and still fit the excess. The tension between the dSph and the Galactic centre excess can therefore be relieved.

\begin{table}\centering
\resizebox{1.0\linewidth}{!}{%
\begin{tabular}{@{}rrrrrrr@{}}
\toprule 
\multicolumn{7}{l}{Best fit points for the real scalar DM case}\\
\midrule
 & $\mdm$ [GeV] & $\sv$ [cm$^3$s$^{-1}$] & $\sixs$ [pb] & $\chi^2_{\mathrm{GCE}}$ (p-value) & $\chi^2_{\mathrm{dSph}}$  & $\chi^2_{\mathrm{\Omega h^2}}$ \\
\midrule
w/ GCE & 49.0 &\num{1.93e-26} & \num{8.52e-11} & 27.74 (0.15)& 71.6 & 0.2 \\
w/o GCE & 173.3 &\num{2.47e-28} & \num{2.22e-10} & -- & 66.7 & \num{1.5e-6} \\
\toprule
\multicolumn{7}{l}{Best fit points for the complex scalar DM case}\\
\midrule
w/ GCE & 42.6 & \num{7.37e-27} & \num{8.30e-11} & 28.2 (0.14) & 67.56 & 0.003 \\
w/o GCE & 2.76 & \num{4.84e-28} &\num{4.82e-4} &  -- & 65.78 & 0.0008\\
\bottomrule
\end{tabular}}
\caption{Best fit points (i.e. minimal $\chi^2$) for both the real and complex scalar DM candidates with and without fitting to the Galactic centre excess. The p-values are calculated only using $\chi^2$ contribution from the Galactic centre excess, under the fairly bold assumption that the test statistic is chi-squared distributed with $24-3=21$ degrees of freedom.\label{t:bfpoints}}
\end{table}

\FloatBarrier

\begin{figure}
\centering
\includegraphics[width=0.90\textwidth]{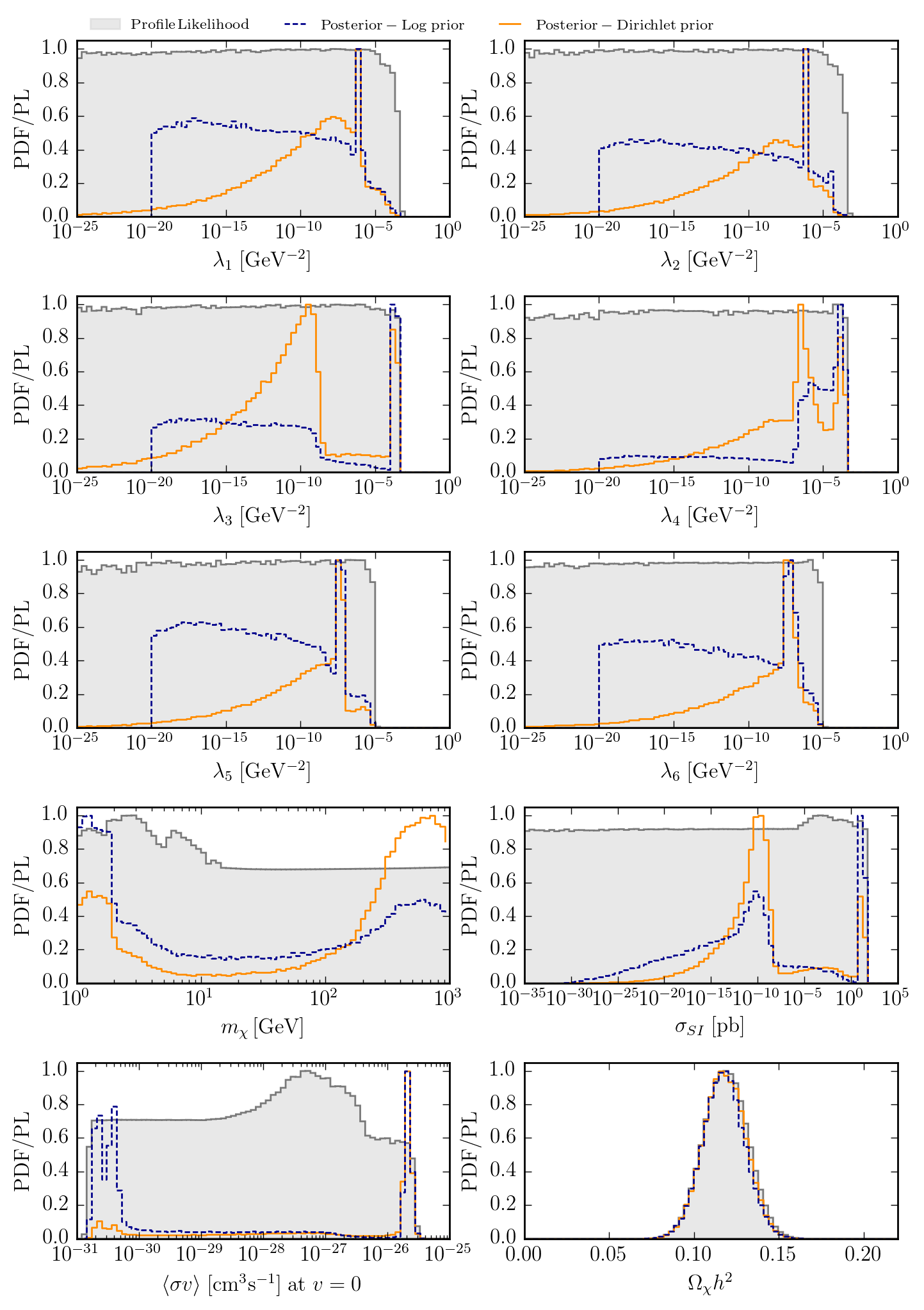}
\caption{1D posterior distributions and profile likelihood of parameters and observables for complex scalar DM.
  The experimental constraints considered are the relic density of DM, limit on spin-independent scattering cross-section from LUX, and annihilation cross-section limits from both the Planck CMB measurement, and the stacked dSph analysis from \Fermi-LAT.
  The different posteriors are using different priors; the Log prior and the Dirichlet prior ($\alpha = 0.1$) as defined in section \ref{s:priors}.
Posteriors and likelihoods are normalised to their peaks. 
 \label{f:C_lim}}
\end{figure}

\FloatBarrier


\section{Conclusions}\label{s:conclusions}

In this article we have presented global scans of combined dark matter--parton EFT operators for both real and complex scalar dark matter, including constraints from cosmology, indirect, and direct detection experiments. We have produced posterior distributions and profile likelihood maps of the model parameter spaces, and thus provided a state of the art, comprehensive, and -- more importantly -- coherent picture of scalar dark matter.

From these distributions we see that, of all current experimental results, the relic density has the greatest impact on the model parameters, as it is an actual measurement and not a limit. All the operator coefficients combined must be just right to provide the correct relic density. The most common configuration, and therefore the most probable, is when a single operator dominates and the others are weak. In contrast with the relic density the null results from LUX and \Fermi-LAT dSph analysis have a somewhat limited impact on the operator coefficients. Instead, they do have effect on the dark matter mass, strongly disfavouring dark matter masses below 100 GeV.

Small regions survive where the dark matter mass is around the charm or the bottom quark mass due to resonance-like effects in the relic density calculation. These regions could very well be excluded by the \Fermi-LAT dSph data if the publicly available likelihood from \Fermi-LAT would be extended to photon energies below $0.5$ GeV.

We also considered the impact of including in our analysis the Galactic centre excess as signal of dark matter. Although standard astrophysical sources might explain the excess, such as unresolved millisecond pulsars, we showed the effect of including it in our likelihood to illustrate the effect of a measurement on the model parameters. The results are dramatic --  a specific mass (40--60 GeV) and operator (R2/C2) are preferred to fit the excess. This region of the parameter space is however in tension with the absence of excess gamma rays from the Milky Way dwarf spheroidals in the real scalar case. So if one believe the Galactic centre excess is due to dark matter, then the quite general real scalar DM model would be excluded. The complex scalar candidate is able to avoid the tension by allowing for lower annihilation cross-section.

Our toolchain is model-independent which means that, in principle, extending these types of combined EFT models -- adding non-parton operators, or  fermionic dark matter -- or looking a more complete models such as simplified models is as easy as writing down the Lagrangian. The main caveat being that adding more degrees of freedom without additional experimental constraints is inadvisable until more informative data become available.

\newpage
{\bf Acknowledgments.} 
 G.B. (P.I.) and S.L. acknowledges support from the European Research Council through the ERC starting grant {\it WIMPs Kairos}.
F.C. and C.W. (P.I.) are part of the VIDI research programme “Probing the Genesis of Dark Matter”, which is financed by the Netherlands Organisation for Scientific Research (NWO).
R. RdA, is supported by the Ram\'on y Cajal program of the Spanish MICINN and also thanks the support of the Spanish 
MICINN's Consolider-Ingenio 2010 Programme 
under the grant MULTIDARK CSD2209-00064, the Invisibles European ITN project (FP7-PEOPLE-2011-ITN, 
PITN-GA-2011-289442-INVISIBLES and the 
``SOM Sabor y origen de la Materia" (FPA2011-29678) and the ``Fenomenologia y Cosmologia de la Fisica mas 
alla del Modelo Estandar e lmplicaciones Experimentales 
en la era del LHC" (FPA2010-17747) MEC projects. Finally also thanks to the Severo Ochoa MINECO project: SEV-2015-0398.
R.T. was partially supported by an EPSRC ``Pathways to Impact" grant. 
The work of TMPT is supported in part by NSF grant PHY-1316792 and by the University of California, Irvine through a Chancellor's Fellowship. 
We gratefully acknowledge the use of the LISA cluster (Amsterdam).

\bibliographystyle{JHEP}
\bibliography{eft}

\end{document}